\documentstyle[12pt]{article}
\begin{document}
\begin{center}
{\bf \large
Raman, infrared and optical spectra of the spin-Peierls compound NaV$_2$O$_5$}
\vspace{10pt}

S.\,A.\,Golubchik$^a$,
M.\,Isobe$^b$,
A.\,N.\,Ivlev$^a$,
B.\,N.\,Mavrin$^a$,
M.\,N.\,Popova$^a$ \footnote{e-mail: popova@isan.troitsk.ru},
A.\,B.\,Sushkov$^a$,
Yu.\,Ueda$^b$, and
A.\,N.\,Vasil'ev$^c$,
\\[5mm]

$^a$Institute of Spectroscopy of RAS, Troitsk, 142092, Russia \\

$^b$Institute for Solid State Physics, The University of Tokyo \\
7--22--1 Roppongi, Minato-ku, Tokyo 106, Japan

$^c$Physics Department, Moscow State University \\
119899 Moscow, Russia \\
\end{center}

Keywords: phonons, Raman spectra, FIR transmission, $d$-electrons. \\
%PACS: 75.30.-m, 78.30.-j \\

%			Abstract

We have measured polarized spectra of Raman scattering, infrared and optical
transmission of NaV$_2$O$_5$ single crystals above the temperature of the
spin-Peierls transition $T_{sp}=35\,K$.
Some of the far-infrared (FIR) phonon lines are strongly asymmetric, due to
the spin-phonon interaction.
In addition to the phonon lines, a broad band was observed in the
$c(aa)\bar{c}$ Raman spectrum and in the ${\bf E}\parallel{\bf a}$ FIR
transmission spectrum. A possible origin of these bands is discussed.
The absorption band at 10000\,cm$^{-1}$ (1.25\,eV) is attributed to vanadium
$d-d$ electronic transitions while the absorption edge above 3\,eV
is supposed to correspond to the onset of charge-transfer transitions.

\newpage
{\bf 1. INTRODUCTION}

NaV$_2$O$_5$  crystallizes in an orthorhombic cell with space group $P2_1mn$
and two formula units in the primitive cell \cite{c1}.
All the atoms occupy twofold positions $2a$ having the symmetry $C_s$.
There are magnetic chains of the V$^{4+}$O$_5$ ($S=1/2$) square pyramids in the
structure isolated by nonmagnetic chains of the V$^{5+}$O$_5$ (S=0) pyramids.
Both binds of chains are oriented along the b-axis.
These chains form layers while sodium ions lie between the layers.
Above 35\,K, the magnetic susceptibility of NaV$_2$O$_5$ is well
described by the
equation for the $S=1/2$  1-D antiferromagnetic (AF) Heisenberg linear chain
with nearest neighbor interaction
${\cal H} = \sum_i \rm
J{\bf S}_{\mit i}{\bf S}_{\mit i+1}$
and exchange constant $J=560\,K$ \cite{c2}.
It has been shown recently \cite{c2}--\cite{c4} that at $T_{sp}=35\,K$
NaV$_2$O$_5$ undergoes the spin-Peierls transition, namely, the spin-singlet
nonmagnetic ground state is formed, simultaneously with a dimerization of the
atoms along the chains.
As a result of the dimerization, an energy gap $\Delta$ opens in the magnetic
excitation spectrum ($\Delta \simeq$9.8\,meV=79\,cm$^{-1}$ in the case
of NaV$_2$O$_5$ \cite{c3}).
The spin-Peierls transition is mediated by the interaction between an
isotropic 1-D $S=1/2$ AF spin system and a 3-D phonon field.
In the weak coupling limit, the BCS-type relation
$2\Delta \simeq3.53\,kT_{sp}$ must be valid \cite{c5}.
All the spin-Peierls compounds studied before obeyed the mentioned relation
rather well.
On the contrary, NaV$_2$O$_5$, the compound with the highest known
$T_{sp}$ demonstrates a strong deviation from this relation
($2\Delta \simeq6.4\,kT_{sp}$ for NaV$_2$O$_5$)
which implies that the spin-phonon coupling is not weak.
Manifestations of this coupling in the far-infrared (FIR)
spectra of NaV$_2$O$_5$
have recently been reported in Ref.\cite{c6}.
To better understand peculiarities of the spin-Peierls transition in
NaV$_2$O$_5$ it is important to know its spin and lattice dynamics and energy
spectrum.

The present paper is devoted, mainly, to a thorough study of the phonon
spectrum of NaV$_2$O$_5$ in the uniform phase (at $T>T_{sp}$)
by means of polarized Raman scattering and far-infrared absorption.
Preliminary data on the electronic spectrum are also presented. \\

{\bf 2. EXPERIMENTAL}

The NaV$_2$O$_5$ single crystals used in this study were obtained as described
in Ref.\cite{c2,c3} and had dimensions of approximately
1$\times$4$\times$0.5\,mm$^3$
along the $a$-, $b$-, and $c$-axes.
The orientation of the crystals and lattice constants
($a=11.318$\,\AA, $b=3.611$\,\AA, $c=4.797$\,\AA)
were determined by X-ray method.
We used, in particular, the samples studied before by the EPR and magnetic
susceptibility measurements \cite{c4}.
The platelets with the thickness of 0.4--40\,$\mu$m along the $c$-axis were
prepared for FIR and optical measurements.

Raman spectra were excited at room temperature by the 514\,nm and 488\,nm
lines of an Ar-ion laser in backscattering geometries, dispersed by a
home-made triple spectrograph and recorded using a multichannel system
consisting of an image intensifier tube with a multichannel plate and a
vidicon.

Polarized transmission measurements were made in the spectral range
50--30000\,cm$^{-1}$ (3\,meV--3.7\,eV) with the resolution ranging from 0.05
to 16\,cm$^{-1}$ at the temperatures 40--300\,K using a Fourier transform
spectrometer BOMEM DA3.002 (for the spectral range 50--20000\,cm$^{-1}$) and
a grating spectrometer Specord M400 (for the spectral range
15000--30000\,cm$^{-1}$). \\

{\bf 3. RESULTS AND DISCUSSION}

A. RAMAN AND INFRARED SPECTRA

The factor group analysis gives the following vibrational representation
for NaV$_2$O$_5$:
\[ \Gamma=16A_1+8A_2+8B_1+16B_2 \]
(Here, the irreducible representations refer to the standard
setting for $Pmn2_1$ space group where the $x$-, $y$-, and $z$-axes are
directed along the $b$-, $c$- and $a$-axes, respectively).
Subtracting three acoustical modes ($A_1+B_1+B_2$), we get 45 optical
modes.
All the optical modes are Raman active. The diagonal components $aa$, $bb$,
$cc$ of the Raman scattering tensor which manifest themselves for the
parallel polarizations ${\bf E_i}\parallel{\bf E_s}$ of the incident and
scattered light are nonzero only for the $A_1$ modes.
In the ${\bf E_i}\perp{\bf E_s}$ geometries another modes are active, namely,
$ab$, $ac$ and $bc$ components of the scattering tensor are nonzero for the
$B_1$, $B_2$ and $A_2$ modes, respectively.
In the FIR absorption, $A_2$ modes are silent, $A_1$, $B_1$, and $B_2$ modes
are active for the polarization of the incident light ${\bf E}\parallel{\bf
a}$, ${\bf E}\parallel{\bf b}$ and ${\bf E}\parallel{\bf c}$, respectively.
In our FIR experiment the wave vector of the incident light was parallel to
the $c$-axis, so we could observe $A_1$ modes (for ${\bf E}\parallel{\bf a}$)
and $B_1$ modes (for ${\bf E}\parallel{\bf b}$).

Figs.\,1--3 show polarized Raman spectra of NaV$_2$O$_5$.
Fig.\,1 presents the spectra for ${\bf E_i}\parallel{\bf E_s}$,
in other words, $A_1$ modes.
Raman spectra turned out to be the most intense in the geometry
$a(cc)\bar{a}$ where longitudinal optical (LO) $A_1$ modes are active.
In the two other geometries, $c(bb)\bar{c}$ and $c(aa)\bar{c}$,
the wave vector of a phonon is
perpendicular to the $a$-axis and, thus, transverse optical (TO) $A_1$ modes
must be active.
Nevertheless, the spectra in these two geometries differ markedly, which
points to the considerable anisotropy of the structure in the directions of
the $a$- and $b$-axes ($b$-axis is parallel to the chains of VO$_5$ pyramids).
Frequencies of all the observed Raman modes are collected in table 1.
Positions of the five observed LO $A_1$ modes are very close to the positions
of some TO $A_1$ modes, indicating small LO--TO splittings and, hence, small
oscillator strengths of these modes.

Besides the phonon lines, we observed a broad band in the $c(aa)\bar{c}$
spectrum, with a maximum near 600\,cm$^{-1}$, extending from 200 to
900\,cm$^{-1}$.
Since this band appears under both 514.5\,nm and 488\,nm excitation
(see fig.\,2), we conclude that it originates from the Raman
scattering process.
According to preliminary data, the broad band near 600\,cm$^{-1}$ is present
down to liquid helium temperatures. Its interpretation is not yet clear.
It follows from the data on the magnetic susceptibility of NaV$_2$O$_5$,
which has a broad maximum around the room temperature, that a short range
order (SRO) and, hence, (damped) magnetic excitations are present
in the V$^{4+}$O$_5$ magnetic chains even at room temperature.
One could expect to observe two-mode magnetic scattering in NaV$_2$O$_5$.
Such a scattering has been found in another spin-Peierls compound CuGeO$_3$
at the temperatures when SRO is preserved \cite{c8}.
However, such a scattering should
have a maximum intensity for parallel polarization of incoming and scattered
light, direction of polarization being along the magnetic chains
\cite{c8} (that is, in $(bb)$ geometry), as is not the case experimentally.
Therefore, it is necessary to consider another interpretation of this band,
for example, two-phonon processes or electronic Raman scattering to a low
lying d-level of the V$^{4+}$ ion. We shall return to this point below.

For crossed polarization of the incident and scattered light,
the spectra turned out to be much weaker than for parallel polarization
(see fig.\,3). From the eight predicted $A_2$ modes we could
find only five in the $a(cb)a$ geometry. They are listed in table~1.
Furthermore, we can not exclude that the lines at 89 and 970\,cm$^{-1}$
appear due to leakage of very intense $(cc)$ components of the Raman tensor.

From the fifteen TO $B_2$ modes we were able to find only seven
in the $b(ac)\bar{b}$ geometry.
In addition, several very weak modes overlap each other in the
region between 390 and 430\,cm$^{-1}$.
Even less number of Raman lines was seen in the $c(ab)\bar{c}$
geometry where transverse $B_1$ modes
should be active. From the expected seven $B_1$ modes we have found only
three. The broad band displayed above for the $c(aa)\bar{c}$ geometry is
present also for the crossed polarization,
though slightly displaced and weak in intensity.

Polarized infrared transmittance spectra of
NaV$_2$O$_5$ in the uniform phase are given in figs.\,4,5.
In fig.\,4 we show the spectral range 50--2500\,cm$^{-1}$
while in fig.\,5 the far-infrared region is displayed in detail.
Peaks in transmittance of a thin crystal are located at the TO phonon
frequencies.
To study modes with different oscillator strengths we used samples of
different thickness.
At room temperature, we were able to detect only three $A_1$
modes and three $B_1$ modes. Below 200~K several more modes
became observable due to the narrowing of their lines.
The TO phonon frequencies found from the polarized FIR transmittance spectra
are listed in table~1.
Some of the phonon lines show a pronounced asymmetry (see fig.\,5).
It should be mentioned here that the thinnest sample (3) used was not uniform
in thickness, so in its case the lineshapes were not correct.
The $B_1$ mode 215\,cm$^{-1}$ and the $A_1$ mode 145\,cm$^{-1}$ have
high frequency tails.
The lowest frequency $A_1$ mode ($\nu$=90\,cm$^{-1}$) has a shape characteristic
for Fano type resonance (the resonance between a descrete level and a continuum
of states~\cite{c7}).
We suppose that asymmetric lineshapes are due to the spin-phonon coupling
that drives the system into a dimerized phase (see also Ref.\cite{c6}).

More thick samples reveal, in addition to the phonon lines, a broad
${\bf E}\parallel{\bf a}$
polarized absorption continuum of a complex shape extending from very low
frequencies (below 50\,cm$^{-1}$) to about 1500\,cm$^{-1}$ (see fig.\,4).
We suppose that at least a part of this continuum arises from the two-mode
magnetic absorption.
It follows from the symmetry consideration that in the case of NaV$_2$O$_5$
such an absorption should have a maximum intensity in the ${\bf
E}\parallel{\bf a}$ polarization \cite{c9,c6}.

The phonon frequencies extracted from our Raman and FIR measurements
are summarized in table~1.
Not all the predicted modes have been found and that does not allow to carry
out calculations of the lattice dynamics and to draw ionic displacement
patterns for the observed modes.
Nevertheless, we present below some considerations concerning a character of
the observed vibrational modes.

The crystal structure of NaV$_2$O$_5$ consists of layers
of the VO$_5$ pyramids
interconnected by their edges or corners \cite{c1,c2}.
Bonds between vanadium and oxygen atoms are predominantly covalent.
Weak interlayer coupling is due to Van-der-Vaals interaction between apical
oxygens of the pyramids in neighboring layers.
Sodium atoms are situated between the VO$_5$ pyramids within a layer and form
ionic bonds with nearest oxygens.
Usually, the intensity of Raman scattering from vibrations of ionically bound
atoms is small as compared to the intensity of scattering from vibrations of
atoms with covalent bonds.
On the contrary, ionic vibrations should be strong in the FIR spectra.
The strongest IR lines 530\,cm$^{-1}$ (A$_1$) and 594\,cm$^{-1}$ ($B_1$) come,
probably, from the Na--O ionic vibrations.

The primitive cell of the NaV$_2$O$_5$ crystal contains only one
layer. Therefore, interlayer vibrations must be absent and all
the observed modes must refer to the intralayer vibrations. The
high frequency modes (in the region of 500\,cm$^{-1}$ and
higher) most probably originate from dominantly oxygen
vibrations in VO$_5$ pyramids while modes below 100\,cm$^{-1}$
dominantly come from vibrations of vanadium atoms. In the
intermediate frequency region both types of atoms participate in
vibrations. According to the structural data \cite{c1}, the
distance between vanadium and apical oxygen is the smallest one
among the V-O distances in the VO$_5$ pyramids. It is equal to
1.65\,\AA\ in V$^{4+}$O$_5$ pyramids and to 1.53\,\AA\ in V$^{5+}$O$_5$
pyramids while the other V-O distances are 1.89--1.98\,\AA\ for the
first kind of pyramids and 1.76--1.98\,\AA\ for the second one. On
that basis we suppose that the highest frequency mode at
970\,cm$^{-1}$ well separated from the other modes originates most
probably from vibrations of the apical oxygens. \\

B. OPTICAL SPECTRA

The region of phonon frequencies is followed by a transparency window up to
5000\,cm$^{-1}$.
We have determined the refractive indexes $n_a$ and $n_b$ of NaV$_2$O$_5$
crystal for ${\bf E}\parallel{\bf a}$ and ${\bf E}\parallel{\bf b}$, from the
interference patterns observed within this window, according to the relation
$\Delta=1/(2dn)$ (here, $\Delta$ is the period of interference, $d$ --- the
thickness of the crystal). We obtained the following average values over the
spectral interval from 2000\,cm$^{-1}$ to 4000\,cm$^{-1}$:
$n_a/n_b=1.38\pm6\%, n_a=3.7\pm0.9, n_b=2.7\pm0.7$.

Above 0.62\,eV (5000\,cm$^{-1}$),
a broad complex absorption band is seen (fig.\,5)
followed by a transmittance window in the blue
region of spectrum and an anisotropic absorption edge in the
ultraviolet (uv).
By the analogy
with vanadium oxydes studied before (see, e.g, \cite{c10}),
we ascribe this
band to $d-d$ transitions of
V$^{4+}$ ions but the uv absorption edge --- to the onset of
charge  transfer transitions. In fig.\,7, we propose a simplified
scheme to account for the observed $d-d$ electronic transitions in
NaV$_2$O$_5$. Fivefold orbitally degenerate energy level of a
$d$-electron in a free V$^{4+}$ ion is split into three $A'$ and two
$A''$ levels by the crystal field of $C_s$ symmetry. There are two
equivalent V$^{4+}$ ions in the unit cell (they are situated in
the neighboring V$^{4+}$--O chains). As a result, each excited
electronic energy level $A'$ or $A''$ of the V$^{4+}$ local symmetry
group $C_s (yz)$ gives rise to $A_1-B_2$ or $A_2-B_1$,
respectively, Davydov doublets of the crystal factor group
$C_{2v}(z)$. Electric dipole optical transitions from the fully
symmetrical ($A_1$) ground crystal state are allowed to the $A_1$,
$B_1$ and $B_2$ levels in
${\bf E}\parallel{\bf a}$, ${\bf E}\parallel{\bf b}$, ${\bf E}\parallel{\bf c}$
polarization, respectively. These selection rules are violated by the
spin-orbit coupling which converts each level into a $E_{1/2}$
Kramers doublet. $E_{1/2} \rightarrow E_{1/2}$ transitions are
allowed for all the polarizations of the incident light and
all the components of the Raman scattering tensor.
However, the transitions obeying the above mentioned
selection rules should dominate.

The Davydov splitting is governed by the interaction between two V$^{4+}$
ions in the neighboring chains, while the width of the total $E(k)$ crystal
energy band depends on the coupling between neighboring V$^{4+}$ ions in the
same chain.
As the interaction within the chain is much stronger than the one between the
chains, the Davydov splitting must be much smaller than the width of the
optical absorption bands.

Possibly, the lowest Davydov pair of levels lies at about 600\,cm$^{-1}$ and
is responsible for a part of the observed continuum in FIR absorption and for
the broad band in Raman spectra, while the next three Davydov pairs lie in
the energy range between 0.74 and 2.1\,eV and lead to the broad absorption
band in this spectral region. \\

Acknowledgement

We are grateful to G.N.Zhizhin for a support. This work was
made possible in part by Grants N95-02-03796-a, N96-02-18114-a and
N96-02-19474-a from the Russian Fund for Fundamental Research.

%\newpage

\begin{table}[h]
\caption{Frequencies of the vibrational modes observed in the Raman and
far-infrared transmission spectra of NaV$_2$O$_5$}

\begin{center}
\begin{tabular}{ c c c c c }
\hline
~ & \multicolumn{3}{|c|}{$\omega_{TO}$\,, cm$^{-1}$} &
$\omega_{LO}$\,, cm$^{-1}$ \\
\cline{2-5}
mode & \multicolumn{2}{|c|}{FIR transmission} &
\multicolumn{1}{c|}{Raman} & Raman \\
\cline{2-3}
~ & \multicolumn{1}{|c|}{40\,K} & 300\,K &
\multicolumn{1}{|c|}{300\,K} & 300\,K \\
\hline
A$_1$ &  90 &  ~  & 89  & 89  \\
~     & 140 & 145 &  ~  &  ~  \\
~     &  ~  &  ~  & 175 & 175 \\
~     &  ~  &  ~  & 215 & 225 \\
~     &  ~  &  ~  & 231 &  ~  \\
~     & 254 & 251 & 254 &  ~  \\
~     &  ~  &  ~  & 300 &  ~  \\
~     &  ~  &  ~  & 415 & 419 \\
~     &  ~  &  ~  & 421 &  ~  \\
~     &  ~  &  ~  & 435 &  ~  \\
~     &  ~  &  ~  & 449 &  ~  \\
~     & 531 & 526 & 531 &  ~  \\
~     &  ~  &  ~  & 970 & 970 \\
~     &  ~  &  ~  &  ~  &  ~  \\
B$_1$ & 178 & 175 & 172 &  ~  \\
~     & 215 &  ~  &  ~  &  ~  \\
~     & 225 &  ~  &  ~  &  ~  \\
~     &  ~  &  ~  & 289 &  ~  \\
~     & 371 & 367 &  ~  &  ~  \\
~     & 594 & 587 &  ~  &  ~  \\
~     &  ~  &  ~  & 678 &  ~  \\
~     &  ~  &  ~  &  ~  &  ~  \\
A$_2$ &  ~  &  ~  &  89 &  ~  \\
~     &  ~  &  ~  & 260 &  ~  \\
~     &  ~  &  ~  & 365 &  ~  \\
~     &  ~  &  ~  & 685 &  ~  \\
~     &  ~  &  ~  & 970 &  ~  \\
~     &  ~  &  ~  &  ~  &  ~  \\
B$_2$ &  ~  &  ~  &  89 &  ~  \\
~     &  ~  &  ~  & 176 &  ~  \\
~     &  ~  &  ~  & 190 &  ~  \\
~     &  ~  &  ~  & 224 &  ~  \\
~     &  ~  &  ~  & 308 &  ~  \\
~     &  ~  &  ~  & 531 &  ~  \\
~     &  ~  &  ~  & 970 &  ~  \\

%\hline
\end{tabular}
\end{center}
\end{table}

%\newpage
~\\
\begin{center} Figure captions
\end{center}

Fig. 1. Room-temperature Raman spectra of NaV$_2$O$_5$ for parallel
polarizations of the incident and scattered light. \\[5mm]
Fig. 2. Room-temperature Raman spectra of NaV$_2$O$_5$ in the
$c(aa)\bar{c}$ geometry at 514.5\,nm and 488\,nm excitation. \\[5mm]
Fig. 3. Room-temperature Raman spectra of NaV$_2$O$_5$ for crossed
polarizations of the incident and scattered light. \\[5mm]
Fig. 4. Polarized IR transmittance spectra of different samples of
NaV$_2$O$_5$. (1) The sample of the thickness $d\simeq13\,\mu$m.
Interference within the sample is seen; $T=300\,K$.
(2) $d\simeq0.8\,\mu$m, the sample on a scotch tape, $T=300\,K$.
(3) $d\simeq0.4\,\mu$m, the sample on a scotch tape; $T=40\,K$.
Arrows indicate phonon frequencies. \\[5mm]
Fig. 5. Polarized FIR transmittance spectra of NaV$_2$O$_5$.
(1) $d\simeq13\,\mu$m, $T_1=300\,K$ (solid line), $T_2=40\,K$ (dashed line);
(2) $d\simeq0.8\,\mu$m, $T=300\,K$;
(3) $d\simeq0.4\,\mu$m, $T=40\,K$. \\[5mm]
Fig. 6. Room-temperature optical absorption spectrum of NaV$_2$O$_5$
single crystal. \\[5mm]
\begin{figure}
\setlength{\unitlength}{1mm}
\begin{picture}(170,200)
\linethickness{1mm}
\put(0,140){\line(1,0){15}}
\put(5,143){\LARGE d}
\put(25,160){\line(1,0){20}}
\put(30,163){\LARGE $^2$E$_g$}
\put(25,120){\line(1,0){20}}
\put(30,123){\LARGE $^2$T$_{2g}$}
%
%--first 3 lines----
%
\put(60,190){\line(1,0){25}}
\put(70,193){\LARGE $^2$B$_1$}
\put(60,150){\line(1,0){25}}
\put(70,153){\LARGE $^2$B$_2$}
\put(60,130){\line(1,0){25}}
\put(70,133){\LARGE $^2$A$_1$}
\put(60, 90){\line(1,0){25}}
\put(70, 93){\LARGE $^2$E}
%
%--following 4 lines end---
%
\put( 90,190){\line(1,0){25}}
\put(100,193){\LARGE $^2$A$'$}
\put( 90,150){\line(1,0){25}}
\put(100,153){\LARGE $^2$A$''$}
\put( 90,130){\line(1,0){25}}
\put(100,133){\LARGE $^2$A$'$}
\put( 90,100){\line(1,0){25}}
\put(100,103){\LARGE $^2$A$'$}
\put( 90, 80){\line(1,0){25}}
\put(100, 83){\LARGE $^2$A$''$}
%
%--following 5 lines end---
%
\put(120,195){\line(1,0){30}}
\put(152,195){\LARGE $^2$B$_2$}
\put(120,185){\line(1,0){30}}
\put(152,185){\LARGE $^2$A$_1$}
\put(120,155){\line(1,0){30}}
\put(152,155){\LARGE $^2$A$_2$}
\put(120,145){\line(1,0){30}}
\put(152,145){\LARGE $^2$B$_1$}
\put(120,135){\line(1,0){30}}
\put(152,135){\LARGE $^2$B$_2$}
\put(120,125){\line(1,0){30}}
\put(152,125){\LARGE $^2$A$_1$}
\put(120,105){\line(1,0){30}}
\put(152,105){\LARGE $^2$A$_1$}
\put(120, 95){\line(1,0){30}}
\put(152, 95){\LARGE $^2$B$_2$}
\put(120, 80){\line(1,0){30}}
\put(152, 80){\LARGE $^2$A$_1$}

\thinlines
\put(15,140){\line(1, 2){10}}
\put(15,140){\line(1,-2){10}}
\put(45,160){\line(1, 2){15}}
\put(45,160){\line(1,-2){15}}
\put(45,120){\line(1, 2){15}}
\put(45,120){\line(1,-2){15}}
\put(85, 90){\line(1, 2){5}}
\put(85, 90){\line(1,-2){5}}
\put(115,190){\line(1, 1){5}}
\put(115,190){\line(1,-1){5}}
\put(115,150){\line(1, 1){5}}
\put(115,150){\line(1,-1){5}}
\put(115,130){\line(1, 1){5}}
\put(115,130){\line(1,-1){5}}
\put(115,100){\line(1, 1){5}}
\put(115,100){\line(1,-1){5}}
\thicklines
\put(126,80){\vector(0,1){105}}
\put(132,80){\vector(0,1){ 65}}
\put(138,80){\vector(0,1){45}}
\put(144,80){\vector(0,1){25}}
\put(127, 85){\LARGE a}
\put(133, 85){\LARGE b}
\put(139, 85){\LARGE a}
\put(145, 85){\LARGE a}
\put(  0, 70){\LARGE free}
\put(  0, 60){\LARGE V$^{4+}$}
\put(  0, 50){\LARGE ion}
\put( 30, 70){\LARGE O$_h$}
\put( 25, 60){\LARGE octa-}
\put( 25, 50){\LARGE hedra}
\put( 67, 70){\LARGE C$_{4v}$}
\put(100, 70){\LARGE C$_s$}
\put( 78, 60){\LARGE square}
\put( 76, 50){\LARGE pyramid}
\put(130, 70){\LARGE C$_{2v}$}
\put(125, 60){\LARGE crystal}
\put(125, 50){\LARGE factor}
\put(125, 40){\LARGE group}
\end{picture}
Fig. 7. Scheme of the $d$-level splitting in a crystal.
Allowed optical transitions are shown only for the
${\bf E}\parallel{\bf a}$ and
${\bf E}\parallel{\bf b}$ polarizations available in our
${\bf k}\parallel{\bf c}$ experimental geometry.
\end{figure}
\end{document}